\documentclass[twocolumn,showpacs,aps,prl,superscriptaddress,floatfix]{revtex4}
\usepackage{graphicx}
\usepackage{verbatim}
\usepackage{dcolumn}
\usepackage{amsmath}
\usepackage{epsfig}
\usepackage{subfigure}

\newcommand{\BaBarYear}      {14}
\newcommand{\BaBarNumber}    {002}
\newcommand{\BaBarType}      {PUB}  
\newcommand{\SLACPubNumber}  {15979}

\input babarsym.tex
\def\ra{\rightarrow}
\def\epem{e^+e^-}
\def\mpmm{\mu^+\mu^-}
\def\lplm{l^+l^-}

\def\figurebox#1#2#3{%
    \def\arg{#3}%
    \ifx\arg\empty
    {\hfill\vbox{\hsize#2\hrule\hbox to #2{\vrule\hfill\vbox to #1{\hsize#2\vfill}\vrule}\hrule}\hfill}%
    \else
    {\hfill\epsfbox{#3}\hfill}%
    \fi}

\begin{document}

\pagestyle{plain}

\begin{flushleft}
\babar-\BaBarType-\BaBarYear/\BaBarNumber \\
SLAC-PUB-\SLACPubNumber\\
arXiv:1406.2980 [hep-ex]\\
\end{flushleft}

\title{{\large \bf Search for a dark photon in $e^+ e^-$ collisions at \babar}}

%
\author{J.~P.~Lees}
\author{V.~Poireau}
\author{V.~Tisserand}
\affiliation{Laboratoire d'Annecy-le-Vieux de Physique des Particules (LAPP), Universit\'e de Savoie, CNRS/IN2P3,  F-74941 Annecy-Le-Vieux, France}
\author{E.~Grauges}
\affiliation{Universitat de Barcelona, Facultat de Fisica, Departament ECM, E-08028 Barcelona, Spain }
\author{A.~Palano$^{ab}$ }
\affiliation{INFN Sezione di Bari$^{a}$; Dipartimento di Fisica, Universit\`a di Bari$^{b}$, I-70126 Bari, Italy }
\author{G.~Eigen}
\author{B.~Stugu}
\affiliation{University of Bergen, Institute of Physics, N-5007 Bergen, Norway }
\author{D.~N.~Brown}
\author{M.~Feng}
\author{L.~T.~Kerth}
\author{Yu.~G.~Kolomensky}
\author{M.~J.~Lee}
\author{G.~Lynch}
\affiliation{Lawrence Berkeley National Laboratory and University of California, Berkeley, California 94720, USA }
% ...
\collaboration{The \babar\ Collaboration}
\noaffiliation

\begin{abstract}
Dark sectors charged under a new Abelian interaction have recently received much attention in the context of dark matter 
models. These models introduce a light new mediator, the so-called dark photon ($A'$), connecting the dark sector 
to the Standard Model. We present a search for a dark photon in the reaction $\epem \rightarrow \gamma A', A' \rightarrow \epem, \mpmm$ 
using 514 $\rm fb^{-1}$ of data collected with the \babar~ detector. We observe no statistically significant deviations from the 
Standard Model predictions, and we set 90\% confidence level upper limits on the mixing strength between the photon and dark photon at the 
level of $10^{-4} - 10^{-3}$ for dark photon masses in the range $0.02 - 10.2\gev$. We further constrain the 
range of the parameter space favored by interpretations of the discrepancy between the calculated 
and measured anomalous magnetic moment of the muon.
\end{abstract}


\maketitle

\setcounter{footnote}{0}

Dark sectors, which introduce new particles neutral under the Standard Model (SM) gauge symmetries, arise in many models of physics 
beyond the Standard Model~\cite{Essig:2013lka}. These particles would only interact feebly with ordinary matter, and could easily have 
escaped detection in past experimental searches. Besides gravity, a few renormalizable interactions provide a portal 
into dark sectors. One of the simplest realizations consists of a dark sector charged under a new gauge group 
$U(1)'$. The corresponding gauge boson, dubbed the dark photon ($A'$), couples to the SM hypercharge via 
kinetic mixing~\cite{Holdom} with a mixing strength $\epsilon$. This results in an effective interaction 
$\epsilon e A'_\mu J^\mu_{EM}$ between the dark photon and the electromagnetic current $J^\mu_{EM}$ after 
electroweak symmetry breaking. This idea has recently received much attention in the context of dark matter models, 
where weakly interacting massive particles reside in a dark sector charged under a new Abelian interaction~\cite{Finkbeiner:2007kk, Pospelov:2007mp, 
ArkaniHamed:2008qn}. Within this framework, dark photons would mediate the annihilation of WIMP particles 
into SM fermions. To accommodate the recent anomalies observed in cosmic rays~\cite{Adriani:2008zr,FermiLAT:2011ab,Aguilar:2013qda}, 
the dark photon mass is constrained to be in the $\mev$ to $\gev$ range.

Low-energy $\epem$ colliders offer an ideal environment to probe low-mass dark sectors~\cite{Batell:2009yf, Essig:2009nc}. Dark 
photons could be produced in association with a photon in $\epem$ collisions, and decay back to SM fermions if other dark sector 
states are kinematically inaccessible. The dark photon width, suppressed by a factor $\epsilon^2$, is expected to be well 
below the experimental resolution. Dark photons could therefore be detected as narrow resonances in radiative 
$\epem \rightarrow \gamma l^+l^-$ ($l=e,\mu$) events. No unambiguous signal for a dark photon has been reported so far, 
and constraints have been set on the mixing strength between the photon and dark photon as a function of the dark photon 
mass~\cite{Blumlein:2011mv, Andreas:2012mt,Endo:2012hp,Babusci:2012cr,Babusci:2014sta,Adlarson:2013eza,Agakishiev:2013fwl,Blumlein:2013cua,
Merkel:2014avp,Abrahamyan:2011gv,Aubert:2009cp,Bjorken:2009mm}. 
Searches for an additional low-mass, dark gauge boson~\cite{arXiv:0908.2821} or dark Higgs boson~\cite{Lees:2012ra} have 
also yielded negative results. 

We report herein a search for dark photons in the reaction $\epem \rightarrow \gamma A', A' \rightarrow \lplm$ ($l=e,\mu$) 
with data recorded by the \babar\ detector~\cite{Bib:Babar,TheBABAR:2013jta}. This 
search is based on 514 fb$^{-1}$ of data collected mostly at the $\Y4S$ resonance, but also at the $\Y3S$ 
and $\Y2S$ peaks, as well as data in the vicinity of these resonances~\cite{Lees:2013rw}. We probe dark photon masses in the 
range $0.02 \gev < m_{A'} < 10.2 \gev$~\cite{units}. To avoid experimental bias, we examine 
the data only after finalizing the analysis strategy. About 5\% of the dataset is used to optimize the selection criteria and 
validate the fitting procedure, and is then discarded from the final data sample.

Simulated signal events are generated by MadGraph~\cite{Alwall:2007st} for 35 different $A'$ mass hypotheses. The background processes 
$\epem \rightarrow \epem (\gamma)$ and $\epem \rightarrow \gamma \gamma (\gamma)$ are simulated using BHWIDE~\cite{Jadach:1995nk} (see below), 
and $\epem \rightarrow \mpmm (\gamma)$ events are generated with KK~\cite{Jadach:2000ir}. Resonance production processes in initial state 
radiation, $\epem \rightarrow \gamma X$ $(X=\jpsi, \psitwos, \Y1S, \Y2S)$, are simulated using a structure 
function technique~\cite{Bib::Struct1,Bib::Struct2}. The detector acceptance and reconstruction efficiencies are determined using a Monte Carlo (MC) 
simulation based on GEANT4 \cite{Bib::Geant}. 

We select events containing two oppositely charged tracks and a single photon having a center-of-mass (CM) energy 
greater than $0.2 \gev$. Additional low-energy photons are allowed if their energies measured in the laboratory frame do not 
exceed $0.2 \gev$. At least one track is required to be identified as an electron, or both tracks as muons, by particle identification 
algorithms. The cosine of the muon helicity angle, defined as the angle between the muon and the CM frame in the $A'$ rest frame, must 
be less than 0.95. To further suppress the contribution from radiative Bhabha events, we also require the cosine of the polar angle (the angle with respect 
to the electron beam axis) of the positron in the CM frame to be larger than $-0.5$, and that of the electron to be less than $0.5$. 
The $\gamma \lplm$ system is then fit, constraining the center-of-mass energy of the candidate to be within the beam energy spread 
and requiring the tracks to originate from the interaction point to within its spread. Finally, we require the $\chi^2$ of the fit 
to be less than 30 (for 8 d.o.f). These criteria are chosen to maximize the signal significance over a broad mass range. 

A large contribution from converted photons produced in $\epem \rightarrow \gamma \gamma, \gamma \rightarrow \epem$ events is 
still present at low $\epem$ invariant mass. A neural network is trained to further reduce this background using the following 
variables: the flight length of the $\epem$ pair in the plane transverse to the beam, and the corresponding flight significance, 
the electron helicity angle, the polar angle of the $\epem$ system, and the angle between the photon and the plane formed by 
the two tracks. We apply a requirement on the neural network output that selects approximately 70\% of the signal in the low-mass 
region, and rejects more than 99.7\% of the photon conversions. The uncertainty associated with this selection criterion, 
estimated from a sample of $\pi^0 \rightarrow \gamma \epem$ decays, is at the level of 2\% at $m_{A'} \sim 20 \mev$, and 
decreases rapidly to negligible levels above $m_{A'} \sim 50 \mev$.

\begin{figure}
\begin{center}
\includegraphics[width=0.5\textwidth]{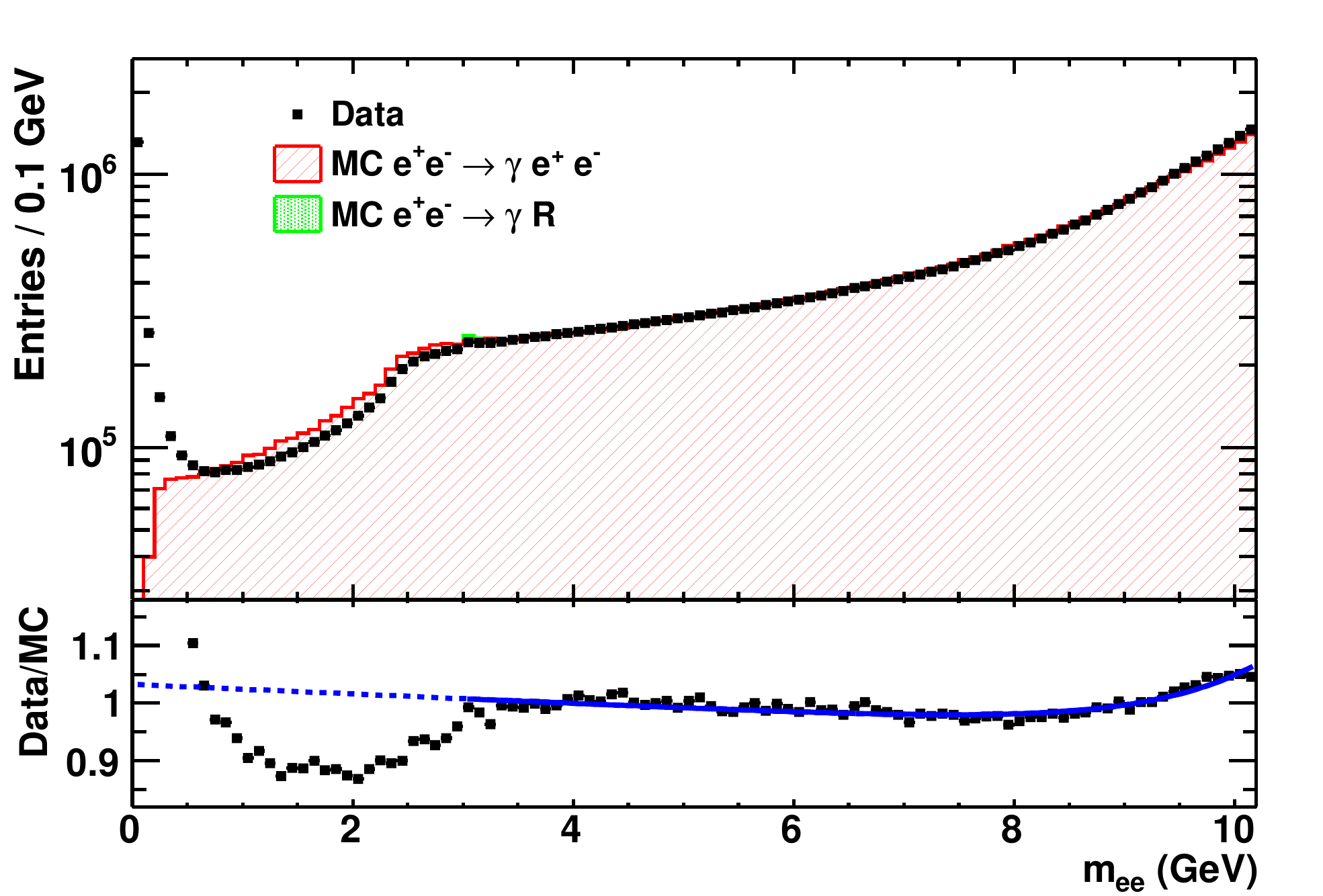}
\includegraphics[width=0.5\textwidth]{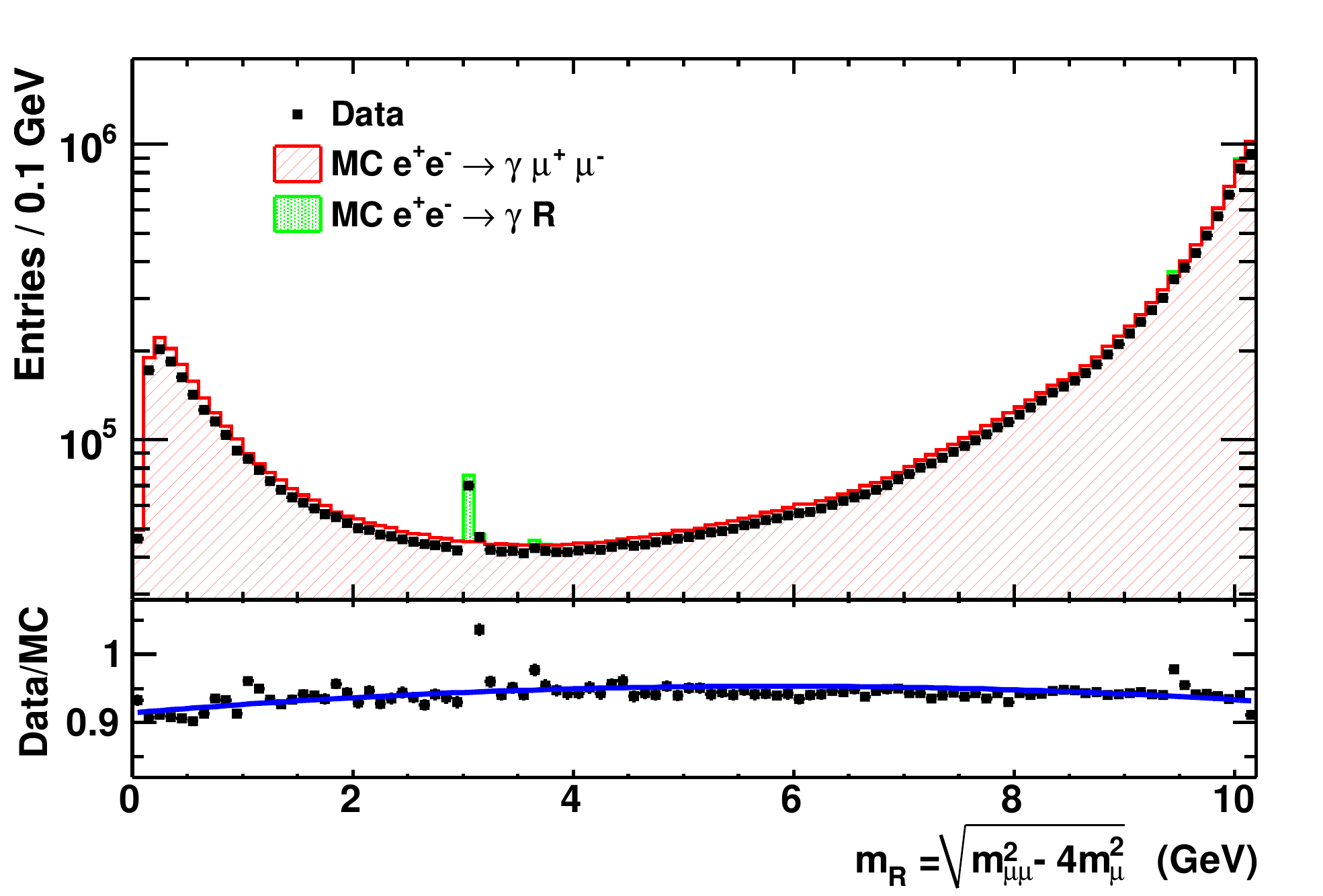}
\caption{Distribution of the final dielectron (top) and reduced dimuon invariant masses (bottom), together with the predictions 
of various simulated SM processes and ISR production of $\jpsi, \psitwos$, $\Y1S$, and $\Y2S$ resonances (collectively 
labeled as R). The fit to the ratio between data and simulated events is described in the text.}
\label{fig1}
\end{center}
\end{figure}

The resulting dielectron and reduced dimuon mass distributions are displayed in Fig.~\ref{fig1}, together with the predictions 
of various simulated SM processes. The reduced dimuon mass, $m_R = \sqrt{m_{\mu\mu}^2 - 4m_\mu^2}$, is easier to model near 
threshold than the dimuon mass. The dielectron (reduced dimuon) mass spectrum is 
dominated by radiative Bhabha (dimuon) production, with smaller peaking contributions from ISR production of 
$\jpsi, \psitwos$, $\Y1S$, and $\Y2S$ resonances. The contribution from $\phi \rightarrow K^+K^-$, where both kaons 
are misidentified as electrons or muons, is found to be negligible. The mass distributions are generally well described 
by the simulation, except in the low $\epem$ mass region, where, as expected, BHWIDE fails to reproduce events in which 
the two leptons are separated by a small angle. Since the signal extraction procedure does not depend on the background 
predictions, this disagreement has little impact on the search.       

The signal selection efficiency, typically 15\% (35\%) for the dielectron (dimuon) channel, is determined from Monte Carlo 
simulation. The difference is mostly due to trigger efficiencies. For electrons, this is lowered in order to suppress 
the rate of radiative Bhabha events. Correction factors to the efficiency, which vary between 0.5\% to 3\%, account for 
the effects of triggers, charged particle identification, and track and photon reconstruction. These are assessed by fitting the 
ratios of the measured and simulated $\epem \rightarrow \epem \gamma$  and $\epem \rightarrow \mpmm \gamma$ differential mass 
distributions, as shown in Fig.~\ref{fig1}. For the dielectron channel, we fit the ratio only in the region $m_{\epem} > 3 \gev$, 
where the simulation is expected to provide reliable predictions, and extrapolate the corrections to the low-mass region. The 
entire mass range is used for the dimuon final state. Half of the corrections are propagated as systematic uncertainties to 
cover statistical variations between neighboring mass points and the uncertainty associated to the extrapolation procedure. 

The signal yield as a function of $m_{A'}$ is extracted by performing a series of independent fits to the dielectron and the reduced 
dimuon mass spectra for each beam energy. The fits are performed in the range $0.02\gev<m_{A'}<10.2 \gev$ 
($0.212\gev<m_{A'}<10.2 \gev$) for the dielectron (dimuon) sample taken near the $\Y4S$ resonance, and up to $9.6 \gev$ and 
$10.0 \gev$ for the dataset collected near the $\Y2S$ and $\Y3S$ resonances, respectively~\cite{units}. We search for a 
dark photon in varying mass steps that correspond to approximately half of the dark photon mass resolution. Each fit is 
performed over an interval at least 20 times broader than the signal resolution at that mass, with the constraint 
$m_{\epem} > 0.015 \gev$ for the dielectron channel. For the purposes of determining the mass steps, the signal resolution 
is estimated by Gaussian fits to several simulated $A'$ samples, and interpolated to all other masses. It varies between 
$1.5$ and $8\mev$. We sample a total of 5704 (5370) mass hypotheses for the dielectron (dimuon) channel. Example of fits 
can be found in the supplemental material~\cite{suppmaterial}. The bias in the fitted values is estimated from a large 
ensemble of pseudo-experiments and found to be negligible.

The likelihood function, described below, contains contributions from signal, radiative dilepton background, and peaking 
background where appropriate. The signal probability density function (pdf) is modeled directly from the signal Monte Carlo mass 
distribution using a non-parametric kernel pdf, and interpolated between the known simulated masses using an algorithm based 
on the cumulative mass distributions~\cite{Read:1999kh}. An uncertainty of 5\%-10\% in this procedure is assessed by taking 
the next-to-closest instead of the closest simulated mass points to interpolate the signal shape. Samples of simulated and 
reconstructed $\epem \rightarrow \gamma \jpsi, \jpsi \rightarrow \lplm$ events indicate that the simulation underestimates 
the signal width by 8\% (4\%) for the dielectron (dimuon) channel. We assume that this difference is independent of the dark 
photon mass, and we increase the signal pdf width by the corresponding amount for all mass hypotheses. We propagate half 
of these correction factors as systematic uncertainties on the fitted signal yields. 

The radiative Bhabha background below $0.1 \gev$ is described by a fourth order polynomial, and elsewhere by a 
third order polynomial. The radiative dimuon background is parametrized by a third order polynomial, constrained 
to pass through the origin for fits in the region below $0.05 \gev$. Peaking contributions from the $\jpsi$, 
$\psitwos$, $\Y1S$, and $\Y2S$ resonances for both final states are included where appropriate. Their shapes are modeled 
as Crystal Ball or Gaussian functions with parameters extracted from fits to the corresponding Monte Carlo samples. Similarly 
to the signal pdf, we increase their width by 8\% (4\%) for the dielectron (dimuon) final states. The interference 
between vector resonances with radiative dilepton production is observed for the $\omega$ and $\phi$ mesons, and is 
fit with the following empirical function:
$$ f(m) = (a+bm+cm^2+dm^3) \left|1-Q\frac{m_{\omega/\phi}\Gamma}{s-m_{\omega/\phi}^2 -im_{\omega/\phi}\Gamma}\right|^2  $$
where $m_{\omega/\phi}$ ($\Gamma$) denotes the mass (width) of the resonance, $Q$ the resonant fraction, and $a,b,c,d$ 
are free parameters. We fix the masses and widths to their nominal values~\cite{pdg}, and let their fractions float. 
We exclude the resonant regions from the search, vetoing ranges of $\pm30 \mev$ around the nominal mass of the 
$\omega$ and $\phi$ resonances, and $\pm 50 \mev$ around the $\jpsi$, $\psitwos$, and $\Upsilon(1S,2S)$ resonances 
(approximately $\pm 5\sigma_R$, where $\sigma_R$ denotes the experimental resolution of the resonances). An alternative 
signal extraction fit, using parametric pdfs for signal~\cite{Aubert:2009cp} and a different background parametrization has been performed for 
the $\mpmm$ channel. The results of both methods are statistically consistent with each other.
The uncertainty on the background modeling is estimated by using an alternative description of the radiative Bhabha and 
dimuon contributions based on a second or fourth order polynomial, depending on the mass hypothesis. This uncertainty is 
almost as large as the statistical uncertainty near the dielectron threshold, and can be as large as 50\% of the statistical 
uncertainty in the vicinity of the $\Upsilon(1S,2S)$ resonances. Outside these regions, the uncertainty varies from a few 
percent at low masses to $\sim 20\%$ of the statistical uncertainty in the high mass region. In addition we propagate half 
of the corrections applied to the signal width, as well as the uncertainties on the $\omega$ and $\phi$ 
masses and widths, as systematic uncertainties on the fitted signal yields. 

\begin{figure}
\begin{center}
\includegraphics[width=0.49\textwidth]{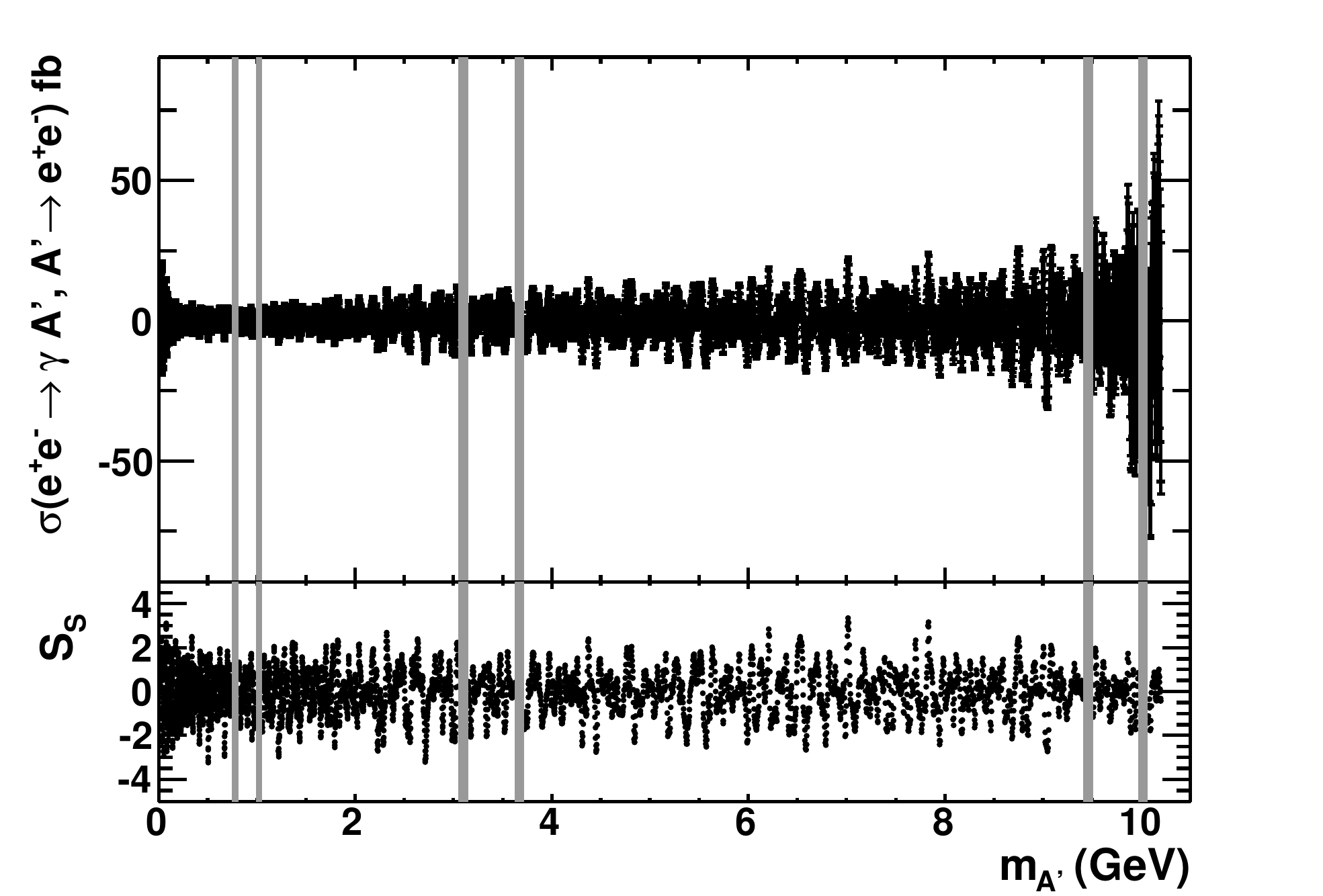}
\includegraphics[width=0.49\textwidth]{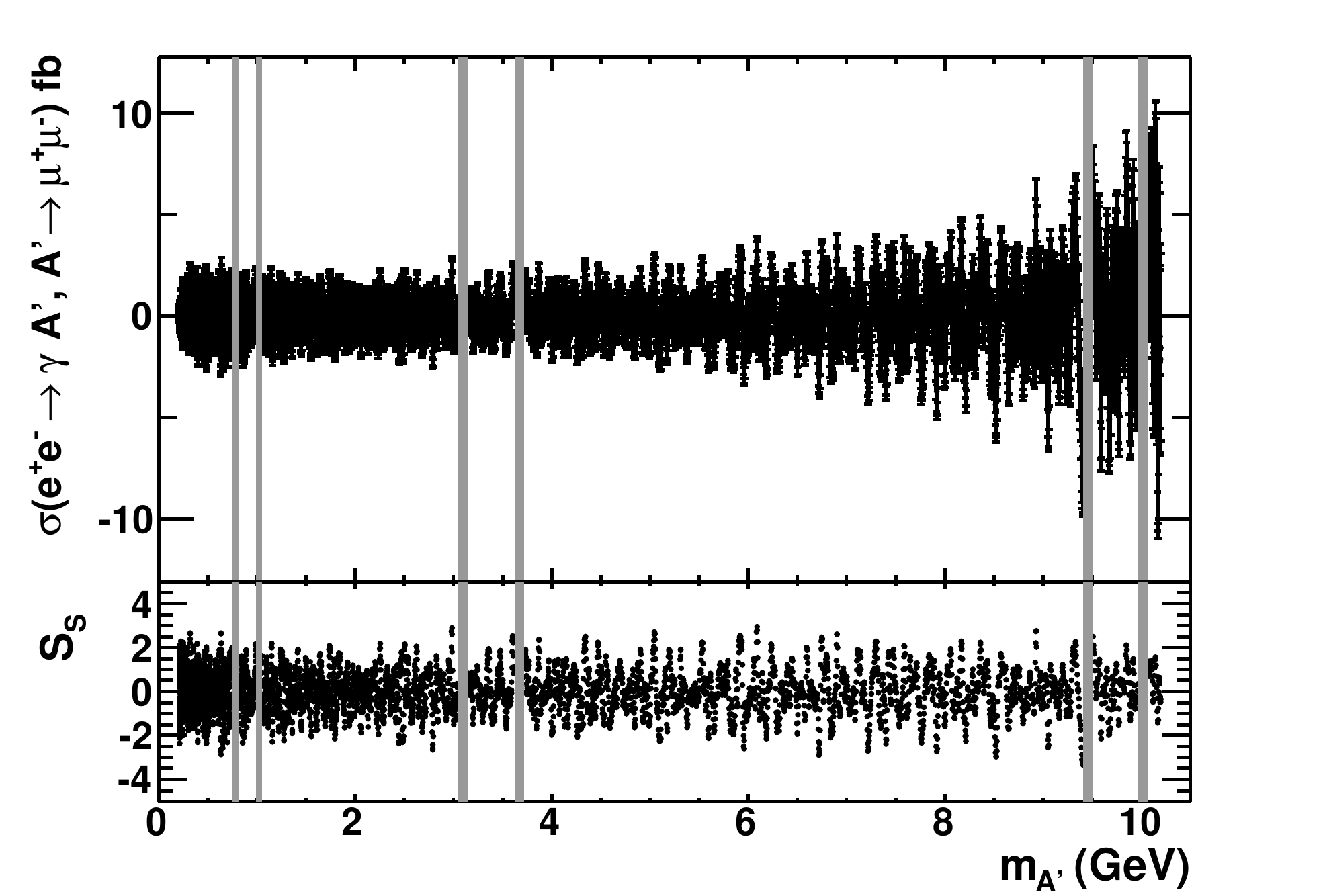}
\caption{The $\epem \rightarrow \gamma A', A' \rightarrow \epem$ (top) and $\epem \rightarrow \gamma A', A' \rightarrow \mpmm$ (bottom) 
cross-sections together with their respective statistical significance ($S_S$) as a function of the dark photon mass. The gray bands 
indicate the mass regions that are excluded from the analysis.}
\label{fig2}
\end{center}
\end{figure}

\begin{figure}
\begin{center}
\includegraphics[width=0.23\textwidth]{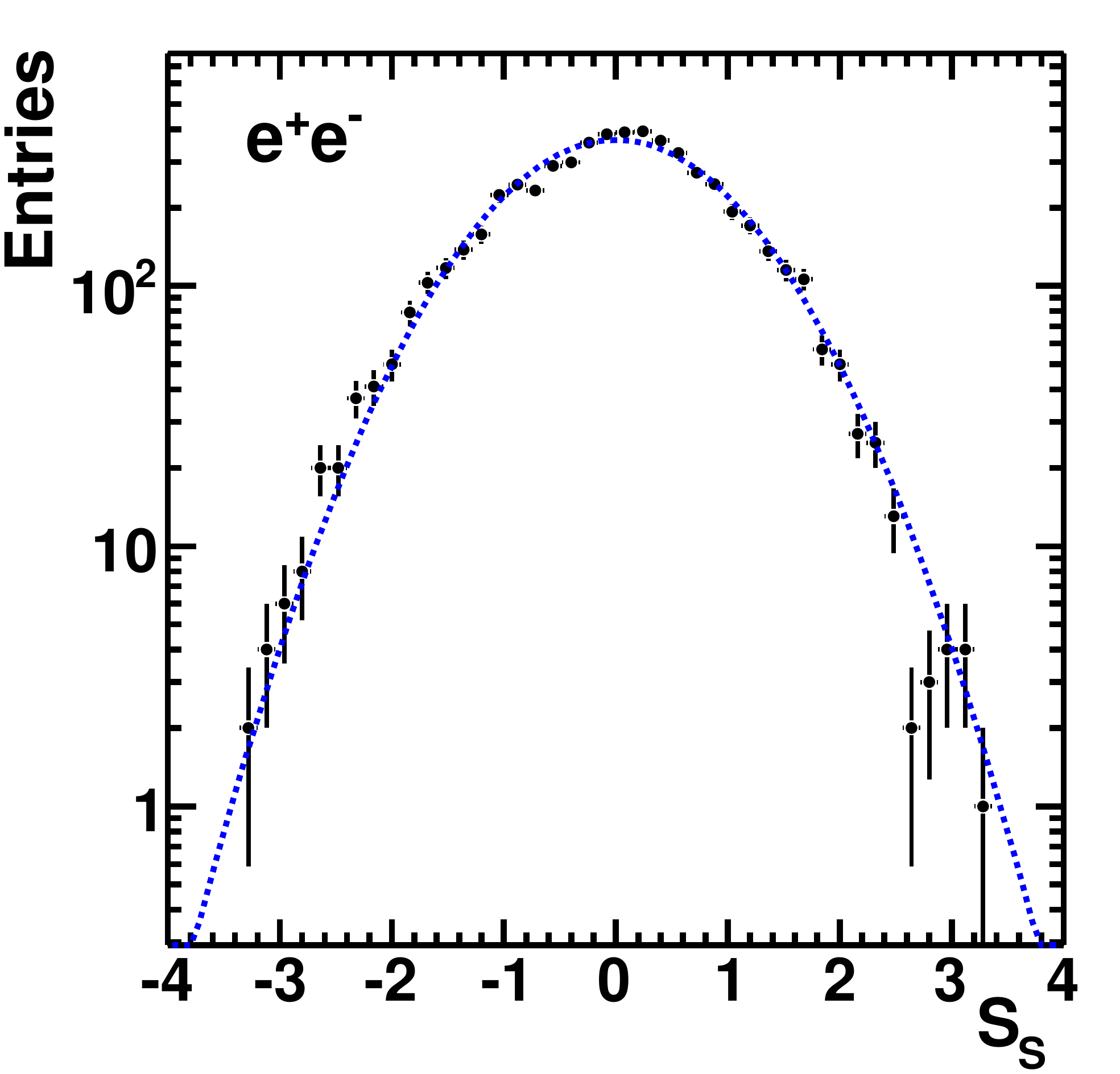}
\includegraphics[width=0.23\textwidth]{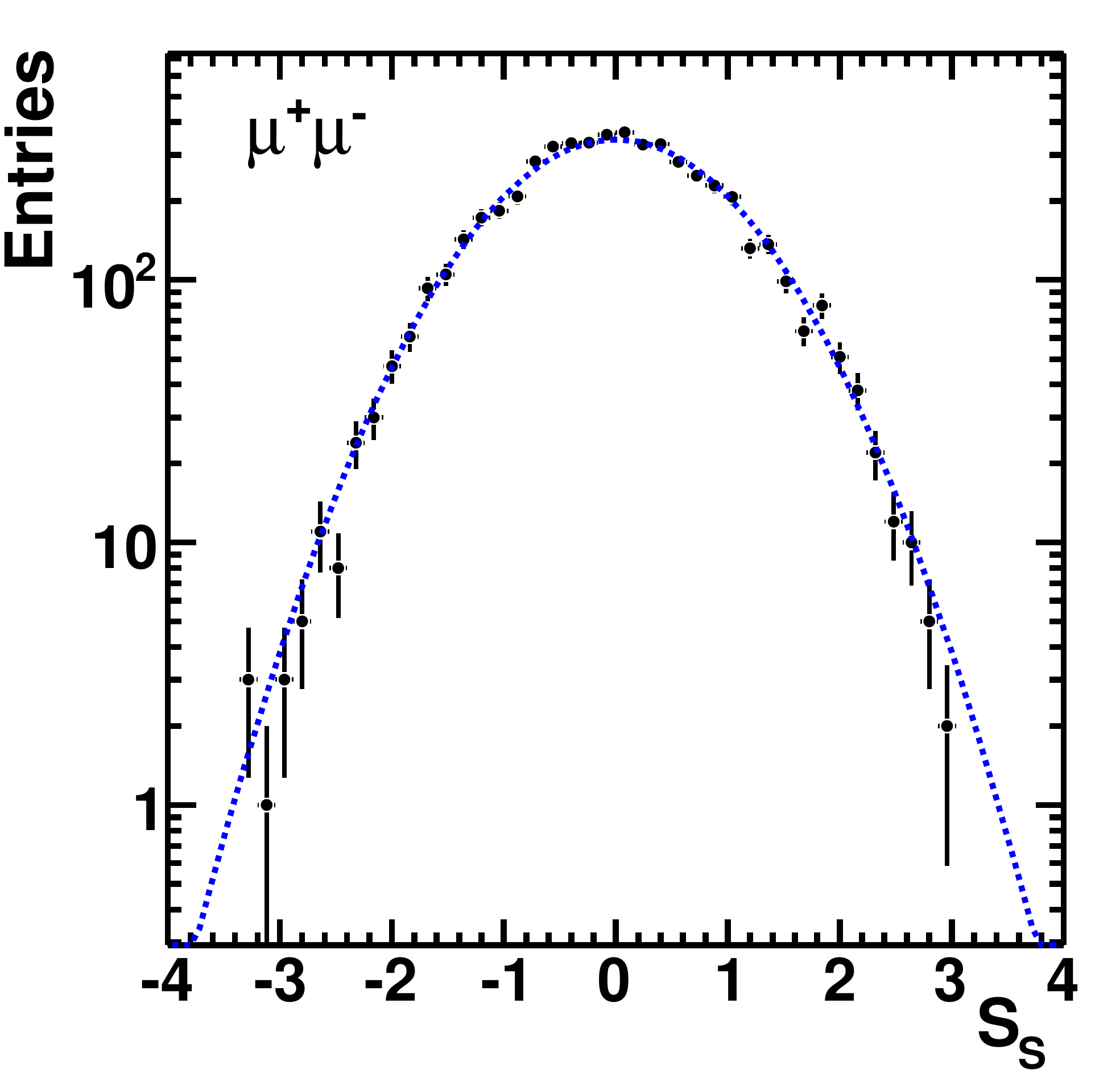}
\caption{Distribution of the statistical significance ($S_S$) from the fits to the dielectron (left) and dimuon (right) 
final states, together with the expected distribution for the null hypothesis (dashed line).}
\label{fig3}
\end{center}
\end{figure}

The $\epem \rightarrow \gamma A', A' \rightarrow \epem$ and $\epem \ra \gamma A', A' \ra \mpmm$ cross-sections as a 
function of the dark photon mass are obtained by combining the signal yields of each data sample, divided by the 
efficiency and luminosity. The cross-sections as a function of $m_{A'}$ are shown in Fig.~\ref{fig2}; the distributions 
of the statistical significances of the fits are displayed in Fig.~\ref{fig3}. The statistical significance of each fit is 
taken as ${\cal S} = \sqrt{2\log{({\cal L / L}_0})}$, where $\cal L$ and ${\cal L}_0$ are the likelihood values for fits with 
a free signal and the pure background hypothesis, respectively. We estimate trial factors by generating a large sample of Monte 
Carlo experiments. The largest local significance is $3.4\sigma$ ($2.9\sigma$), observed near $m_{A'} = 7.02 \gev$ ($6.09 \gev$) for 
the dielectron (dimuon) final state. Including trial factors, the corresponding p-value is 0.57 (0.94), consistent 
with the null hypothesis.

\begin{figure}
\begin{center}
\includegraphics[width=0.5\textwidth]{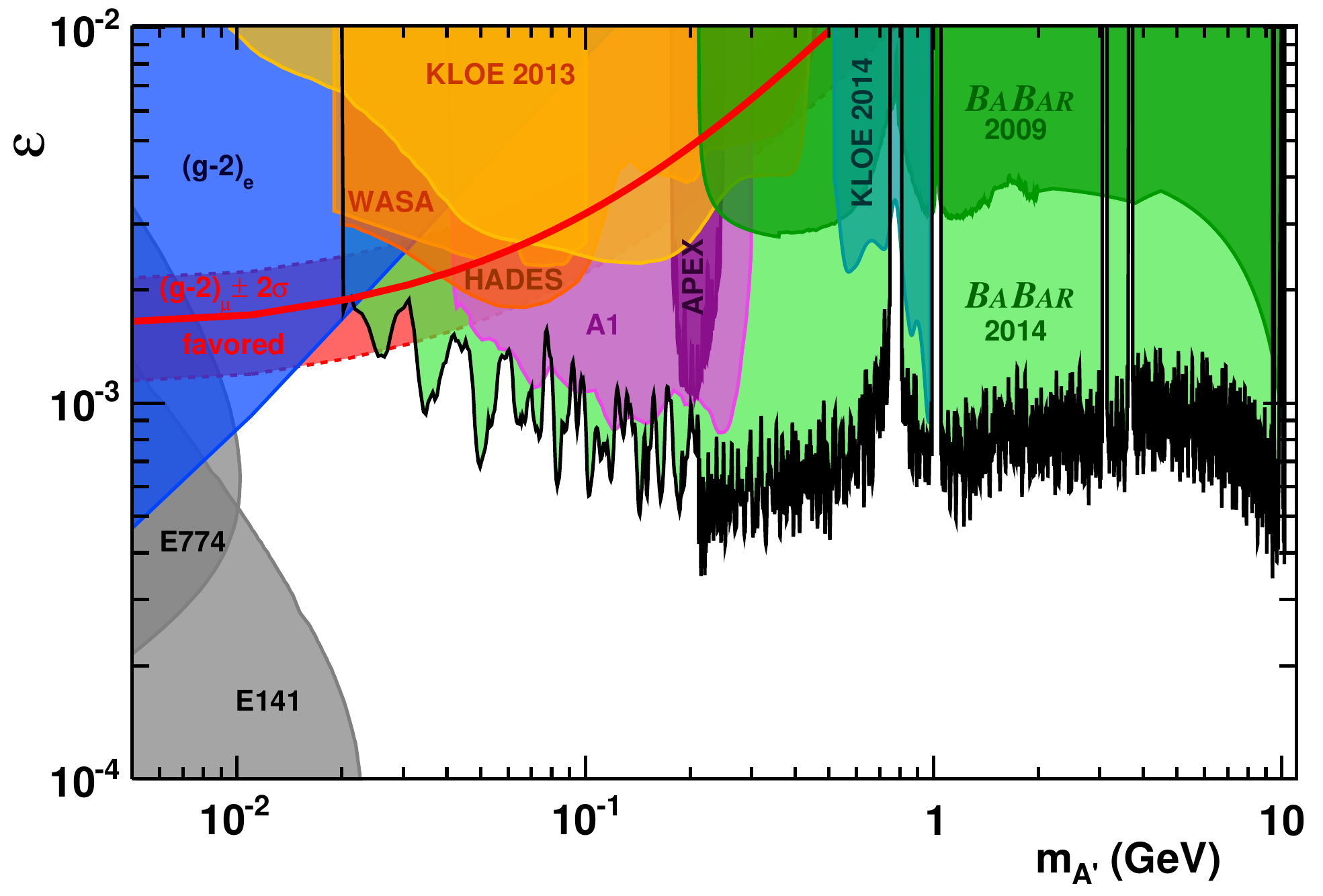}
\caption{Upper limit (90\% CL) on the mixing strength $\epsilon$ as a function of the dark photon mass. The values required to explain 
the discrepancy between the calculated and measured anomalous magnetic moment of the muon~\protect\cite{Pospelov:2008zw} are displayed as a red line.}
\label{fig4}
\end{center}
\end{figure}

We extract the $\epem \rightarrow \gamma A'$ cross-section for each final state using the expected dark photon 
branching fractions $A' \rightarrow \lplm$ from Ref.~\cite{Batell:2009yf}, and combine the results into a single 
measurement. The uncertainties on the dark photon branching fractions (0.1\%-4\%), the luminosity (0.6\%), and the limited 
Monte Carlo statistics (0.5-4\%) are propagated as systematic uncertainties. We derive 90\% confidence 
level (CL) Bayesian upper limits on the $\epem \rightarrow \gamma A'$ cross-section, assuming 
a flat prior for the cross-section. The limits are typically at the level of ${\cal O}(1-10) \rm \, fb$. These results are 
finally translated into 90\% CL upper limits on the mixing strength between the photon and dark photon as a function of 
the dark photon mass~\cite{Essig:2009nc}. The results are displayed in Fig.~\ref{fig4}. The average correlation between 
neighboring points is around 90\%. Bounds at the level of $10^{-4} - 10^{-3}$ for $0.02\gev<m_{A'}<10.2 \gev$ are set, 
significantly improving previous constraints derived from beam-dump experiments~\cite{Blumlein:2011mv,Andreas:2012mt,Blumlein:2013cua}, the electron anomalous 
magnetic moment~\cite{Endo:2012hp}, KLOE~\cite{Babusci:2012cr,Babusci:2014sta}, WASA-at-COSY~\cite{Adlarson:2013eza}, HADES~\cite{Agakishiev:2013fwl}, 
A1 at MAMI~\cite{Merkel:2014avp}, and the test run from APEX~\cite{Abrahamyan:2011gv}. These results also supersede and extend 
the constraints based on a search for a light \CP-odd Higgs boson at \babar~\cite{Aubert:2009cp,Bjorken:2009mm} with a smaller 
dataset. No signal consistent with the excess reported by the HyperCP experiment close to $214 \mev$ is observed~\cite{Park:2005eka, Pospelov:2008zw}. 
We further constrain the range of the parameter space favored by interpretations of the discrepancy between 
the calculated and measured anomalous magnetic moment of the muon~\cite{Pospelov:2008zw}. The remaining mass region of allowed 
parameters, $ 15 \mev \lesssim m_{A'} \lesssim 30 \mev$ will be probed by several planned experiments in the near 
future (see for example ref.~\cite{Essig:2013lka} for a discussion).
 
In conclusion, we have performed a search for dark photon production in the range $0.02\gev<m_{A'}<10.2 \gev$. No significant 
signal has been observed and upper limits on the mixing strength $\epsilon$ at the level of $10^{-4} - 10^{-3}$ have been set. These 
bounds significantly improve the current constraints, and exclude almost all of the remaining region of the parameter 
space favored by the discrepancy between the calculated and measured anomalous magnetic moment of the muon.

\section{Acknowledgments}
\label{sec:Acknowledgments}
The authors wish to thank Rouven Essig and Sarah Andreas for providing us the constraints derived from existing 
experiments. We also thank Rouven Essig, Philip Schuster and Natalia Toro for useful discussions and for 
providing us with their MadGraph code to simulate dark photon processes.
We are grateful for the excellent luminosity and machine conditions
provided by our \pep2\ colleagues, 
and for the substantial dedicated effort from
the computing organizations that support \babar.
The collaborating institutions wish to thank 
SLAC for its support and kind hospitality. 
This work is supported by
DOE
and NSF (USA),
NSERC (Canada),
CEA and
CNRS-IN2P3
(France),
BMBF and DFG
(Germany),
INFN (Italy),
FOM (The Netherlands),
NFR (Norway),
MES (Russia),
MICIIN (Spain),
STFC (United Kingdom). 
Individuals have received support from the
Marie Curie EIF (European Union),
the A.~P.~Sloan Foundation (USA)
and the Binational Science Foundation (USA-Israel).


\begin{thebibliography}{9}
  
\bibitem{Essig:2013lka} 
  See for example R.~Essig, J.~A.~Jaros, W.~Wester, P.~H.~Adrian, S.~Andreas, T.~Averett, O.~Baker and B.~Batell {\it et al.}, 
  arXiv:1311.0029 [hep-ph], and references therein.

\bibitem{Holdom} B. Holdom, Phys.\ Lett.\ B {\bf 166}, 196 (1986).

\bibitem{Finkbeiner:2007kk} 
  D.~P.~Finkbeiner and N.~Weiner, Phys.\ Rev.\ D {\bf 76}, 083519 (2007).

\bibitem{Pospelov:2007mp}
  M.~Pospelov, A.~Ritz, and M.~B.~Voloshin, Phys.\ Lett.\  B {\bf 662}, 53 (2008).

\bibitem{ArkaniHamed:2008qn}
  N.~Arkani-Hamed, D.~P.~Finkbeiner, T.~R.~Slatyer, and N.~Weiner, Phys.\ Rev.\  D {\bf 79}, 015014 (2009).
  
\bibitem{Adriani:2008zr} 
  O.~Adriani {\it et al.}  [PAMELA Collaboration], Nature {\bf 458}, 607 (2009).

\bibitem{FermiLAT:2011ab} 
  M.~Ackermann {\it et al.}  [Fermi LAT Collaboration], Phys.\ Rev.\ Lett.\  {\bf 108}, 011103 (2012).

\bibitem{Aguilar:2013qda} 
  M.~Aguilar {\it et al.}  [AMS Collaboration], Phys.\ Rev.\ Lett.\  {\bf 110}, 141102 (2013).
  
\bibitem{Batell:2009yf}
  B.~Batell, M.~Pospelov, and A.~Ritz, Phys.\ Rev.\  D {\bf 79}, 115008 (2009).

\bibitem{Essig:2009nc}
  R.~Essig, P.~Schuster, and N.~Toro, Phys.\ Rev.\  D {\bf 80}, 015003 (2009).

\bibitem{Blumlein:2011mv}
  J.~Blumlein and J.~Brunner, Phys.\ Lett.\ B {\bf 701} 155 (2011).

\bibitem{Andreas:2012mt} 
  S.~Andreas, C.~Niebuhr and A.~Ringwald, Phys.\ Rev.\ D {\bf 86}, 095019 (2012).
  
\bibitem{Endo:2012hp} 
  M.~Endo, K.~Hamaguchi and G.~Mishima, Phys.\ Rev.\ D {\bf 86}, 095029 (2012).

\bibitem{Babusci:2012cr} 
  D.~Babusci {\it et al.}  [KLOE-2 Collaboration], Phys.\ Lett.\ B {\bf 720}, 111 (2013).

\bibitem{Babusci:2014sta} 
  D.~Babusci {\it et al.}  [KLOE-2 Collaboration], Phys.\ Lett.\ B {\bf 736}, 459 (2014).

\bibitem{Adlarson:2013eza} 
  P.~Adlarson {\it et al.}  [WASA-at-COSY Collaboration], Phys.\ Lett.\ B {\bf 726}, 187 (2013).

\bibitem{Agakishiev:2013fwl} 
  G.~Agakishiev {\it et al.}  [HADES Collaboration], Phys.\ Lett.\ B {\bf 731}, 265 (2014).

\bibitem{Blumlein:2013cua}
  J.~Blmlein and J.~Brunner, Phys.\ Lett.\ B {\bf 731} 320 (2014).

\bibitem{Merkel:2014avp} 
  H.~Merkel, P.~Achenbach, C.~A.~Gayoso, T.~Beranek, J.~Bericic, J.~C.~Bernauer, R.~Boehm and D.~Bosnar {\it et al.},
  Phys.\ Rev.\ Lett.\  {\bf 112}, 221802 (2014).

\bibitem{Abrahamyan:2011gv} 
  S.~Abrahamyan {\it et al.}  [APEX Collaboration], Phys.\ Rev.\ Lett.\  {\bf 107}, 191804 (2011).
  
\bibitem{Aubert:2009cp} 
  B.~Aubert {\it et al.}  [\babar\ Collaboration], Phys.\ Rev.\ Lett.\  {\bf 103}, 081803 (2009).

\bibitem{Bjorken:2009mm} 
 J.~D.~Bjorken, R.~Essig, P.~Schuster and N.~Toro, Phys.\ Rev.\ D {\bf 80}, 075018 (2009).

\bibitem{arXiv:0908.2821} 
  B.~Aubert {\it et al.} [\babar\ Collaboration], arXiv:0908.2821 [hep-ex].

\bibitem{Lees:2012ra} 
  J.~P.~Lees {\it et al.}  [\babar\ Collaboration], Phys.\ Rev.\ Lett.\  {\bf 108}, 211801 (2012).

\bibitem{Bib:Babar}
  B.~Aubert {\it et al.} [\babar\ Collaboration], Nucl.\ Instrum.\ Meth.\ A {\bf 479}, 1 (2002).

\bibitem{TheBABAR:2013jta} 
  B.~Aubert {\it et al.}  [\babar\ Collaboration], Nucl.\ Instrum.\ Meth.\ A {\bf 729}, 615 (2013).

\bibitem{Lees:2013rw} 
   J.~P.~Lees {\it et al.}  [\babar\ Collaboration], Nucl.\ Instrum.\ Meth.\ A {\bf 726}, 203 (2013).

\bibitem{units} Natural units ($\hbar=c=1$) are used throughout this paper.

\bibitem{Alwall:2007st} 
  J.~Alwall, P.~Demin, S.~de Visscher, R.~Frederix, M.~Herquet, F.~Maltoni, T.~Plehn, and D.~L.~Rainwater {\it et al.},
  JHEP {\bf 0709}, 028 (2007).

\bibitem{Jadach:1995nk} 
  S.~Jadach, W.~Placzek and B.~F.~L.~Ward, Phys.\ Lett.\ B {\bf 390}, 298 (1997).

\bibitem{Jadach:2000ir} 
  S.~Jadach, B.~F.~L.~Ward and Z.~Was,  Phys.\ Rev.\ D {\bf 63}, 113009 (2001).

\bibitem{Bib::Struct1} 
  A.~B.~Arbuzov {\em et al.}, J. High Energy Phys. {\bf 9710}, 001 (1997).

\bibitem{Bib::Struct2} 
  M.~Caffo, H.~Czy\.z, and E.~Remiddi, Nuovo Cim. {\bf A110}, 515  (1997); 
  Phys. Lett. {\bf B327}, 369 (1994).

\bibitem{Bib::Geant}
  S.~Agostinelli {\it et al.} (GEANT4 Collab.), Nucl.\ Instrum.\ Methods Phys.\ Res., Sect. A {\bf 506}, 250 (2003).
  
\bibitem{suppmaterial}
  See Supplemental Material at [url] for example of fits to the data and additional plots on the limits on 
  $\epem \rightarrow \gamma A', A' \rightarrow \lplm$ cross-sections.

\bibitem{Read:1999kh} 
  A.~L.~Read, Nucl.\ Instrum.\ Meth.\ A {\bf 425}, 357 (1999).

\bibitem{pdg}   
  J.~Beringer {\it et al.} (Particle Data Group), Phys.\ Rev.\  {\bf D 86}, 010001 (2012) 
  and 2013 partial update for the 2014 edition. 
   
\bibitem{Park:2005eka} 
  H.~Park {\it et al.}  [HyperCP Collaboration], Phys.\ Rev.\ Lett.\  {\bf 94}, 021801 (2005).

\bibitem{Pospelov:2008zw}
  M.~Pospelov, Phys.\ Rev.\  D {\bf 80}, 095002 (2009).


\end{thebibliography}
\end{document}